\documentclass[letterpaper, 11 pt, conference]{ieeeconf}  

\usepackage{cite}
\usepackage{graphicx}
\usepackage{url}
\usepackage{subcaption}
\usepackage{amsmath}
\usepackage{amsfonts}

\renewcommand{\baselinestretch}{1.1} 
\IEEEoverridecommandlockouts                              

\title{\LARGE \bf  On the Traffic Impacts of Optimally Controlled Connected and Automated Vehicles}

\author{Liuhui Zhao, {\itshape{Member, IEEE}}, Andreas A. Malikopoulos, {\itshape{Senior Member, IEEE}}, \\
Jackeline Rios-Torres, {\itshape{Member, IEEE}}
\thanks{This research was supported in part by ARPAE's NEXTCAR program under the award number DE-AR0000796 and in part by U.S. Department of Energy (DOE) Vehicle Technologies Office (VTO) under the Systems and Modeling for Accelerated Research in Transportation (SMART) Mobility Laboratory Consortium, an initiative of the Energy Efficient Mobility Systems (EEMS) Program.}
\thanks{Liuhui Zhao and Andreas A. Malikopoulos are with the Department of Mechanical Engineering, University of Delaware, Newark, DE 19716 USA (emails: \tt\small{lhzhao@udel.edu, andreas@udel.edu).}}
\thanks{Jackeline Rios-Torres is with the Energy and Transportation Science Division, Oak Ridge National Laboratory, Oak Ridge TN 37932 USA USA (phone: 865-946-1542; e-mail: 
       \tt\small{riostorresj@ornl.gov)}.}   
}

\begin{document}

\maketitle
\thispagestyle{empty}

\begin{abstract}
The implementation of connected and automated vehicle (CAV) technologies enables a novel computational framework for real-time control actions aimed at optimizing energy consumption and associated benefits. Several research efforts reported in the literature to date have proposed decentralized control algorithms to coordinate CAVs in various traffic scenarios, e.g., highway on-ramps, intersections, and roundabouts. However, the impact of optimally coordinating CAVs on the performance of a transportation network  has not been thoroughly analyzed yet. In this paper, we apply a decentralized optimal control framework in a transportation network and compare its performance to a baseline scenario consisting of human-driven vehicles. We show that introducing of CAVs yields radically improved roadway capacity and network performance. 
\end{abstract}

\indent

\section{Introduction}
Urban intersections, merging roadways, roundabouts, and speed reduction zones along with the driver responses to various disturbances \cite{Malikopoulos2013} are the primary sources of bottlenecks that contribute to traffic congestion \cite{Margiotta2011}. Connectivity and automation in vehicles provide the most intriguing opportunity for enabling users to better monitor transportation network conditions and make better operating decisions. However, investigating the impact of connected and automated vehicles (CAVs) on a transportation network and related implications on mobility and safety has been of great concern in recent studies \cite{Guanetti2018}.

Several research efforts have been reported in the literature proposing different approaches on coordinating CAVs at different transportation segments, e.g., intersections, roundabouts, merging roadways, speed reduction zones, with the intention to improve traffic flow.  In 2004, Dresner and Stone \cite{Dresner2004} proposed the use of the reservation scheme to control a single intersection of two roads with vehicles traveling with similar speed on a single direction on each road. Some approaches have focused on coordinating vehicles at intersections to improve the travel flow  \cite{Zohdy2012,Yan2009,Li2006,Zhu2015a,Wu2014,kim2014}. Lee and Park \cite{Lee2012} studied coordination of CAVs in an intersection where a phase conflict map is used to remove stop-and-go traffic signals. The approach was extended \cite{Lee2013} to an urban corridor that consisted of multiple intersections.  A detailed discussion of the research efforts in this area can be found in \cite{Malikopoulos2016a, Guanetti2018}.

Most recently, a decentralized optimal control framework was established for coordinating online CAVs in different transportation scenarios. A closed-form, analytical solution was presented and tested in merging roadways \cite{Rios-Torres2015, Ntousakis:2016aa, Rios-Torres2, Malikopoulos2018b}, intersections  \cite{Zhang2016a,Malikopoulos2017, Zhang:2017aa, Mahbub2019ACC}, roundabouts \cite{Malikopoulos2018a}, and speed reduction zones \cite{Malikopoulos2018c}. Rios-Torres and Malikopoulos \cite{Rios2018} discussed the traffic and fuel consumption impacts of partial penetration of CAVs for a highway on-ramp scenario. 

In previous work \cite{Zhao2018}, we discussed the potential benefits of optimally coordinating CAVs in a corridor such that stop-and-go driving is eliminated. However, it is still not clear how a group of optimally controlled CAVs changes the traffic patterns in a  network. In this paper, we explore the impact of the optimal decentralized control framework the we developed in previous work \cite{Rios-Torres2,Malikopoulos2017,Malikopoulos2018c}  for optimally controlling CAVs in a corridor under different traffic conditions, and we identify the benefits as well as the limitations. 

The structure of the paper is organized as follows. In Section II, we introduce the optimal control framework. In Section III, we present the simulation platform and the traffic scenarios that we consider. We provide the evaluation results in Section IV and concluding remarks in Section V.

\section{Optimal Control Framework}
\begin{figure}[ht]
\centering
\includegraphics[width=0.48\textwidth]{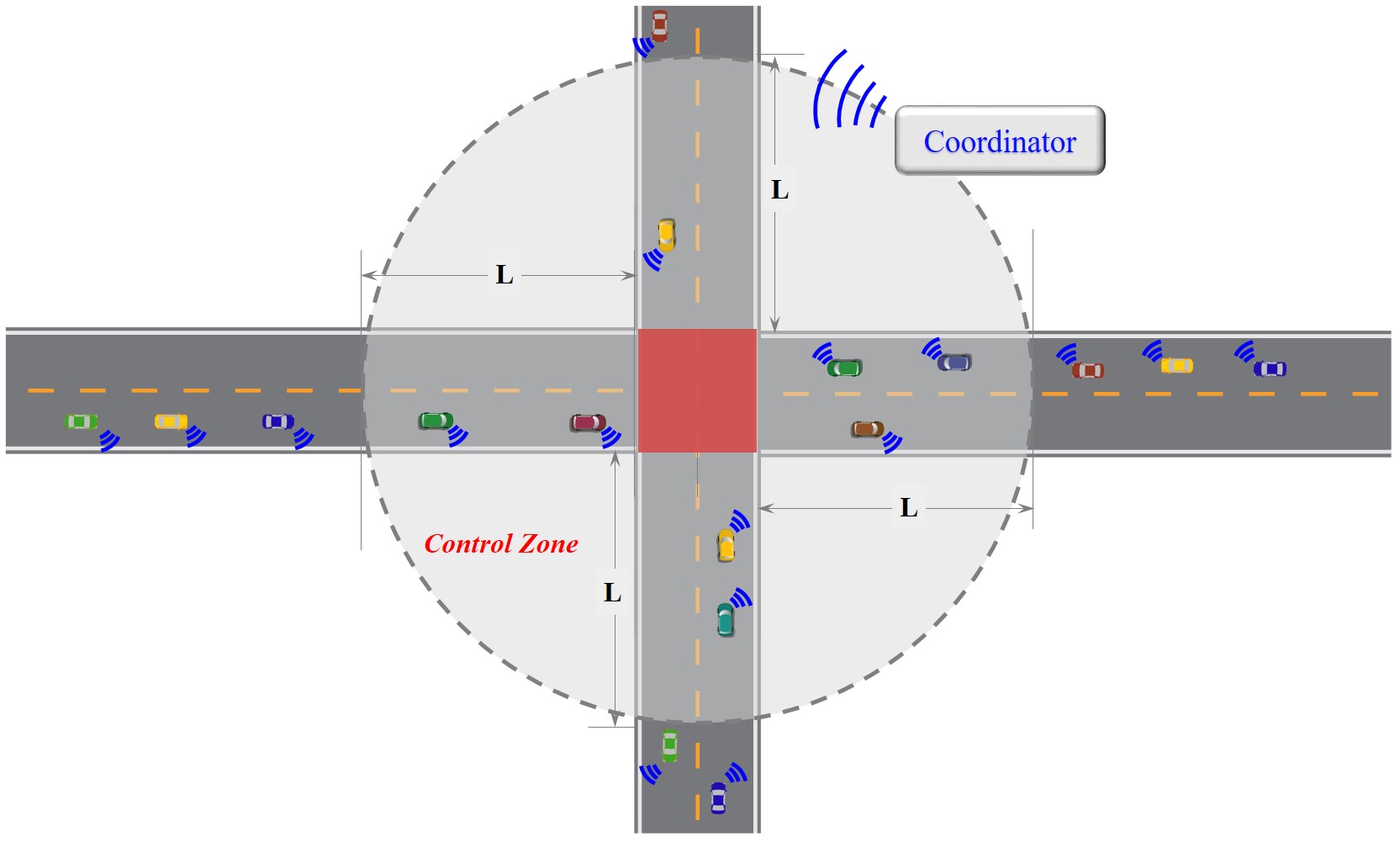}
\caption{An intersection with optimally controlled CAVs.}
\label{fig:control_zone}
\end{figure}

We briefly review the model presented in \cite{Malikopoulos2017}. We consider a \emph{conflict zone} where traffic from two different roadways may cause potential lateral collisions, indicated in red in Fig.~\ref{fig:control_zone}. Before the entry of the conflict zone, there is a \emph{control zone} and a coordinator that can communicate with the vehicles traveling inside the control zone. Note that the coordinator is not involved in any decision of the vehicles. The length of the control zone is $L$ and it is adjustable.

We consider a number of CAVs $N(t)\in \mathbb{N}$, where $ t\in \mathbb{R}^+$ is the time, entering the control zone. Let $\mathcal{N}(t)={1,\dots,N(t)}$ be a queue of vehicles in the control zone. The dynamics of each vehicle $i\in \mathcal{N}(t)$ can be represented by a state equation
\begin{equation} \label{eq:state}
\dot{x}_i(t) = f(t, x_i, u_i), ~ x_i(t_i^0) = x_i^0,
\end{equation}
where $x_i(t), u_i(t)$ are the state and control input of the vehicle, $t_i^0$ is the time that vehicle $i$ enters the control zone, and $x_i^0$ is the initial value of the state. For simplicity, we model each vehicle as a double integrator, i.e., $\dot{p}_i = v_i(t)$ and $\dot{v}_i = u_i(t)$, where $p_i(t) \in \mathcal{P}_i, v_i(t) \in \mathcal{V}_i$, and $u_i(t) \in \mathcal{U}_i$ denote the position, speed, and acceleration/deceleration (control input) of each vehicle $i$. Let $x_i(t)=[p_i(t)~ v_i(t)]^T$ denotes the state of each vehicle $i$, with initial value $x_i^0=[0 ~ v_i^0]^T$, taking values in the state space $\mathcal{X}_i=\mathcal{P}_i\times  \mathcal{V}_i$. The sets $\mathcal{P}_i, \mathcal{V}_i$, and $\mathcal{U}_i, i\in \mathcal{N}(t)$, are complete and totally bounded subsets of $\mathbb{R}$. The state space $\mathcal{X}_i$ for each vehicle $i$ is closed with respect to the induced topology on $\mathcal{P}_i\times  \mathcal{V}_i$ and thus, it is compact.

To ensure that the control input and vehicle speed are within a given admissible range, the following constraints are imposed
\begin{equation}\label{eq:constraints}
\begin{aligned} 
u_{min} &\leq u_i(t) \leq u_{max},~ \text{and} \\
0 &\leq v_{min} \leq v_i(t) \leq v_{max}, ~ \forall t \in [t_i^0, ~ t_i^f],
\end{aligned} 
\end{equation}
where $u_{min}, u_{max}$ are the minimum deceleration and maximum acceleration respectively, and $v_{min}, v_{max}$ are the minimum and maximum speed limits respectively. $t_i^0$ is the time that vehicle $i$ enters the control zone and $t_i^f$ is the time that vehicle $i$ exits the conflict zone.

To ensure the absence of rear-end collision of two consecutive vehicles traveling on the same lane, the position of the preceding vehicle should be greater than, or equal to the position of the following vehicle plus a predefined safe distance $\delta_i(t)$, where $\delta_i(t)$ is proportional to the speed of vehicle $i$, $v_i(t)$. Thus, we impose the rear-end safety constraint
\begin{equation} \label{eq:safety}
s_i(t)=p_k(t)-p_i(t)\geq \delta_i(t), \forall t \in  [t_i^0, ~ t_i^f], 
\end{equation}
where vehicle $k$ is immediately ahead of $i$ on the same lane. 

We consider the problem of minimizing the control input (acceleration/deceleration) for each vehicle $i\in\mathcal{N}(t)$  from the time $t_i^0$ that the vehicle enters the control zone until the time $t_i^f$ that it exits the control zone. Thus, we formulate the following optimization problem for each vehicle in the queue $\mathcal{N}(t)$
\begin{gather} \label{eq:min}
\min_{u_i}\frac{1}{2}\int_{t_i^{0}}^{t_i^{f}} u_i^2(t)dt, ~\forall i \in \mathcal{N}(t), \\
\text{subject to}: (\ref{eq:state}), (\ref{eq:constraints}),  (\ref{eq:safety}), \nonumber\\
p_{i}(t_i^{0})=p_{i}^{0}\text{, }v_{i}(t_i^{0})=v_{i}^{0}\text{, }p_{i}(t_i^{f})=p_i^f,\nonumber\\
\text{and given }t_i^{0}\text{, }t_i^{f}.\nonumber
\end{gather}

The closed-form, analytical solution of this problem has been reported in \cite{Malikopoulos2017, Rios-Torres2, Malikopoulos2018c}.

\section{Simulation Evaluation}
\subsection{Simulation Environment}
\begin{figure}[!ht]
\centering
\includegraphics[width=0.48\textwidth]{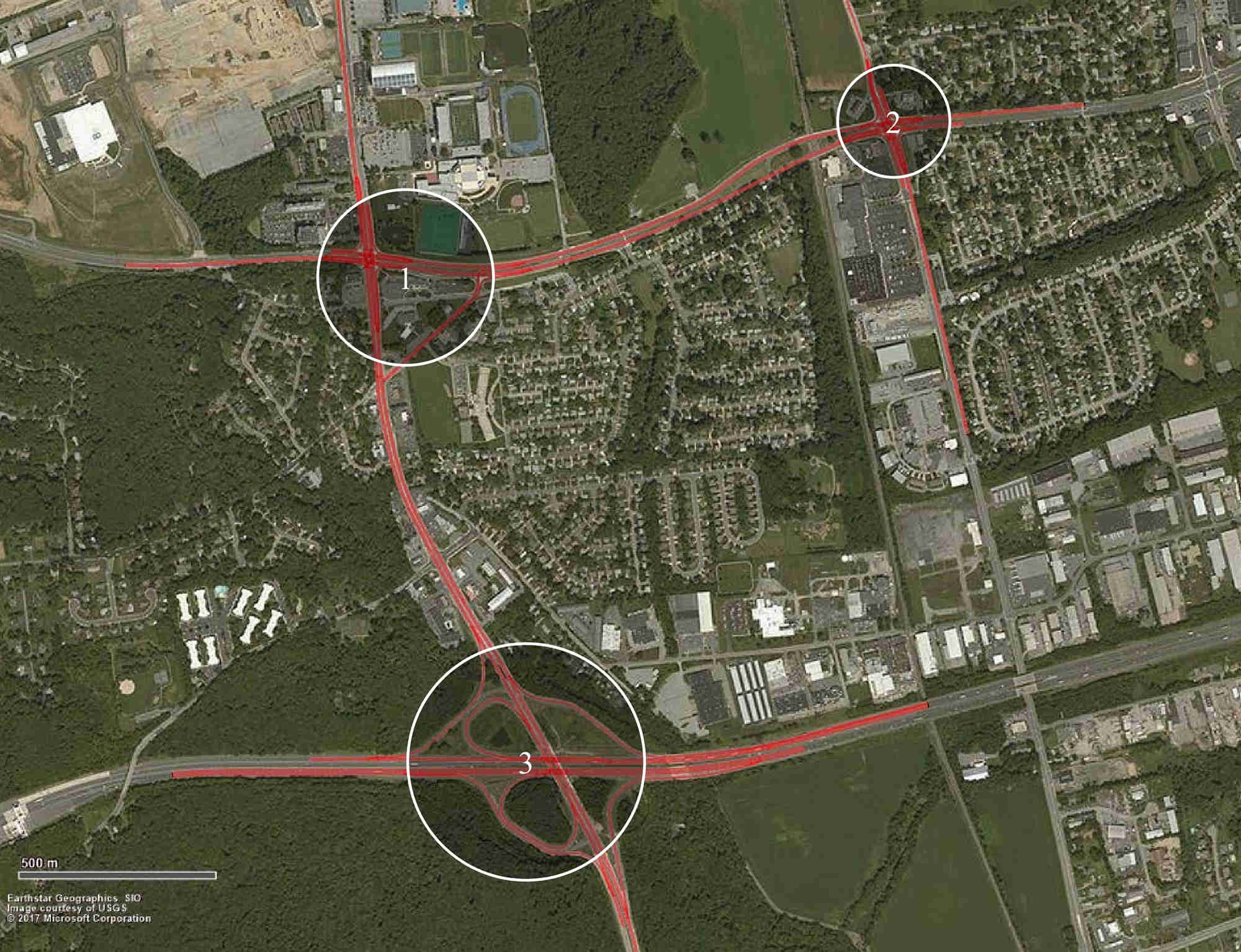}
\caption{Area considered for the case study.}
\label{fig:network}
\end{figure}

To compare network performance between conventional human-driven vehicles, uncontrolled automated vehicles (AVs), and optimally controlled CAVs, we create a simulation network in PTV VISSIM environment in Newark, Delaware area, which includes two signalized intersections (\#1 and \#2 in Fig. \ref{fig:network}, two-lane each direction with dedicated left-turn and right-turn lanes at both intersections, one-lane on southbound at intersection \#2), an unsignalized intersection between \#1 and \#2, and a major interchange (\#3 in Fig. \ref{fig:network}, four-lane each direction, two separated ez-pass only lanes on westbound that we do not consider in the simulation network). In terms of network size, the east-west road segment is approximately 2 $km$, and the north-south segment is about 2.2 $km$. We obtain hourly traffic  information from 2017 annual average daily traffic report published by Delaware Department of Transportation \cite{DelDOT}, and simulate the morning peak hour (i.e., 8:00AM - 9:00AM) traffic. The traffic signal timings are optimized based on current network traffic condition (75 $s$ cycle length for intersection 2 and 90 $s$ cycle length for intersection 1) for the base case. 

For simulating human-driven vehicles, we apply the Wiedemann car following model \cite{wiedemann1974} that is adopted in VISSIM, relating the minimum safe distance as a function of standstill distance and time headway \cite{ptv2018}. For this study, the default 1.5 $m$ standstill distance and 1.2 $s$ time headway are set for simulating the human drivers' car following behavior. Acknowledging that more AVs are on road nowadays, VISSIM adapts its car following model to simulate the AV's driving behavior through adjusting the variation in vehicle acceleration and following distance accordingly. Besides these adjustment, we adopt 0.9 $s$ time headway to reflect the capability of an AV to follow a leading vehicle closer.

As discussed in the previous section, under optimal control operation, each CAV inside the control zone determines its own optimal acceleration profile to drive through the conflict zone smoothly without stop-and-go behavior. Once a vehicle exits an intersection or a merging area, the control algorithm is deactivated such that the vehicle follows any leading vehicle based on the Wiedemann car following model. We consider a control length of 150 $m$ for the coordination of CAVs. In correspondence with different driving behavior between human-driven vehicles and AVs, we apply two different time headway settings (i.e., 0.9 $s$ and 1.2 $s$) in our optimal control algorithm for fair comparison with human-driven vehicles and AVs.


\section{Evaluation Results}
\subsection{Base case with existing traffic demand}
In one hour simulation period, there is approximately 12,000 vehicles entering the network, half of which is served by the interchange. Traffic data is collected every 60 $s$ and vehicle trajectory data is collected every 1 $s$. 
Note that there are both local and highway segments in the network that we considered for the case study. We collect individual vehicle trajectories to analyze vehicle behavior on different routes and try to understand how the optimal control framework affects different groups of traffic. 

We categorize the routes into four groups based on the origin and destination road types: a) local traffic, i.e., both origin and destination roads are arterial; b) highway traffic, i.e., both origin and destination roads are highway; c) local to highway; and d) highway to local. The average travel distances for the four groups are 1.7 $km$, 1.5 $km$, 1.7 $km$, and 2.0 $km$, respectively. We consider the following scenarios: (1) headway 1.2 $s$ with fixed time of the traffic signal controller; (2) headway 0.9 $s$ with fixed time of the traffic signal controller; (3) headway 1.2 $s$, 150 $m$ long control zone, and disabled traffic signals; and (4) headway 0.9 $s$, 150 $m$ long control zone, and disabled traffic signals.


With the optimal control framework, the following distances are increased for all four traffic groups. There are two major reasons: (1) vehicle movements inside the control zones are optimally controlled such that the desired time headway could be guaranteed at each conflict zone; and (2) vehicle stopping at each conflict zone is eliminated as a result of coordination (shown in Table \ref{tab:veh_performance}). Furthermore, looking into vehicle lane change behavior, we find that the average number of lane change is reduced under scenarios 3 and 4. Especially, almost a half of the total lane changes is eliminated for highway traffic, e.g., an average of 0.34 under scenario 1 vs. 0.18 under scenario 3 in Table \ref{tab:veh_performance}. For human-driven vehicles or uncontrolled AVs, it is usually the case that mainline vehicles (especially those on the travel lanes closest to the ramp) would prefer to make lane changes near ramps to avoid interruptions from ramp vehicles. Whereas, under optimal control, all CAVs coordinate with each other such that lane change is not necessary for mainline vehicles. As a result, the number of lane changes is significantly reduced in the network. Through eliminating stop-and-go driving and reducing the number of lane changes, the optimal control framework not only improves traffic mobility, i.e., reduced average travel time as shown in Table \ref{tab:veh_performance}, but also improves safety in the network.


\begin{table}[h]
	\centering
		\caption{Trip performance measurements under different scenarios.}
		\includegraphics[width=.46\textwidth]{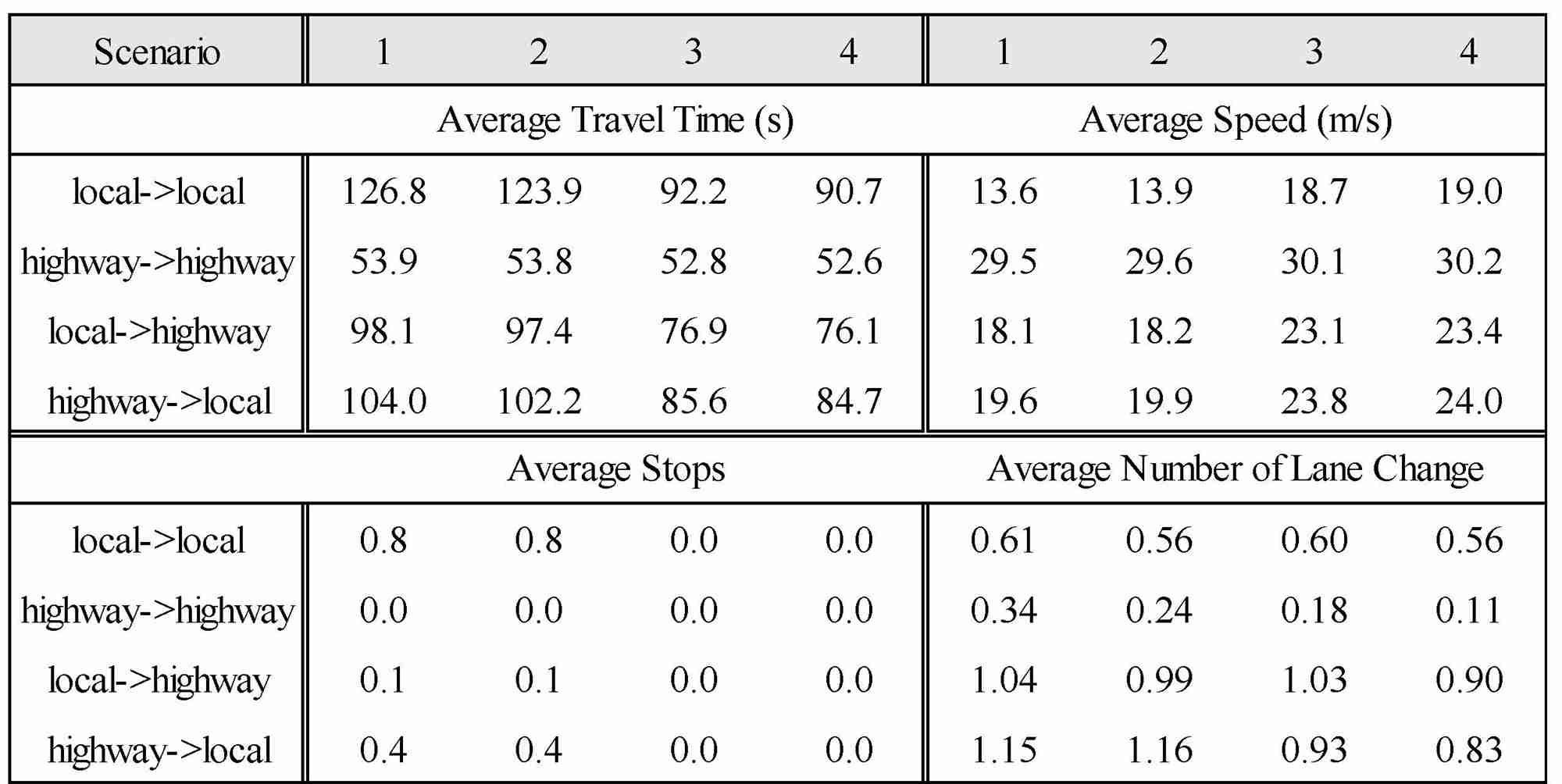} 
	\label{tab:veh_performance}%
\end{table}

\subsection{Sensitivity analysis with varying traffic demand levels}
\begin{figure}[!h]
\centering
\includegraphics[width=.45\textwidth]{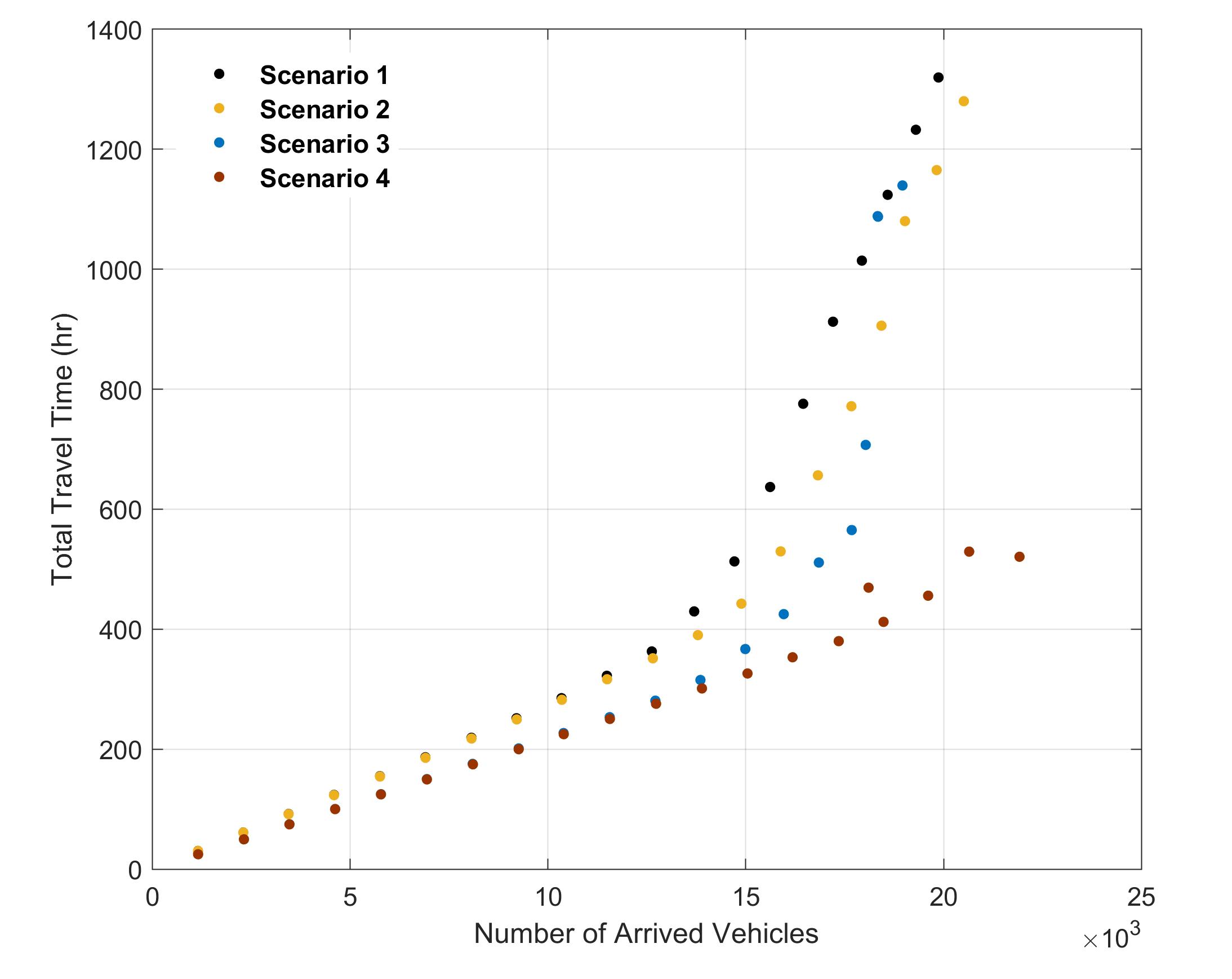}
\caption{Total number of arrived vehicles vs. total travel time in the network.}
\label{fig:throughput}
\end{figure}

To evaluate the effectiveness of the optimal control framework under different traffic conditions, we conduct simulation for a set of traffic demand levels, i.e., 10\% - 200\% of the existing traffic demand, and collect traffic data on the road segments between intersections 1 and 2, and between intersection 1 and the interchange. 
In Fig. \ref{fig:throughput}, we plot the number of arrived vehicles versus total travel time by these vehicles for all the traffic demand levels under four different scenarios, with points representing the data under different demand levels. We find that, under scenarios 1 and 2, the network operates well with any traffic demand level below 150\%, where the total travel time linearly increases with the number of arrived vehicles. When the traffic demand level is above 150\%, the time spent to serve an additional 10\% of vehicles increases exponentially.

\begin{figure*}[!h]
\centering
\begin{subfigure}[b]{0.45\textwidth}
\includegraphics[width=\textwidth]{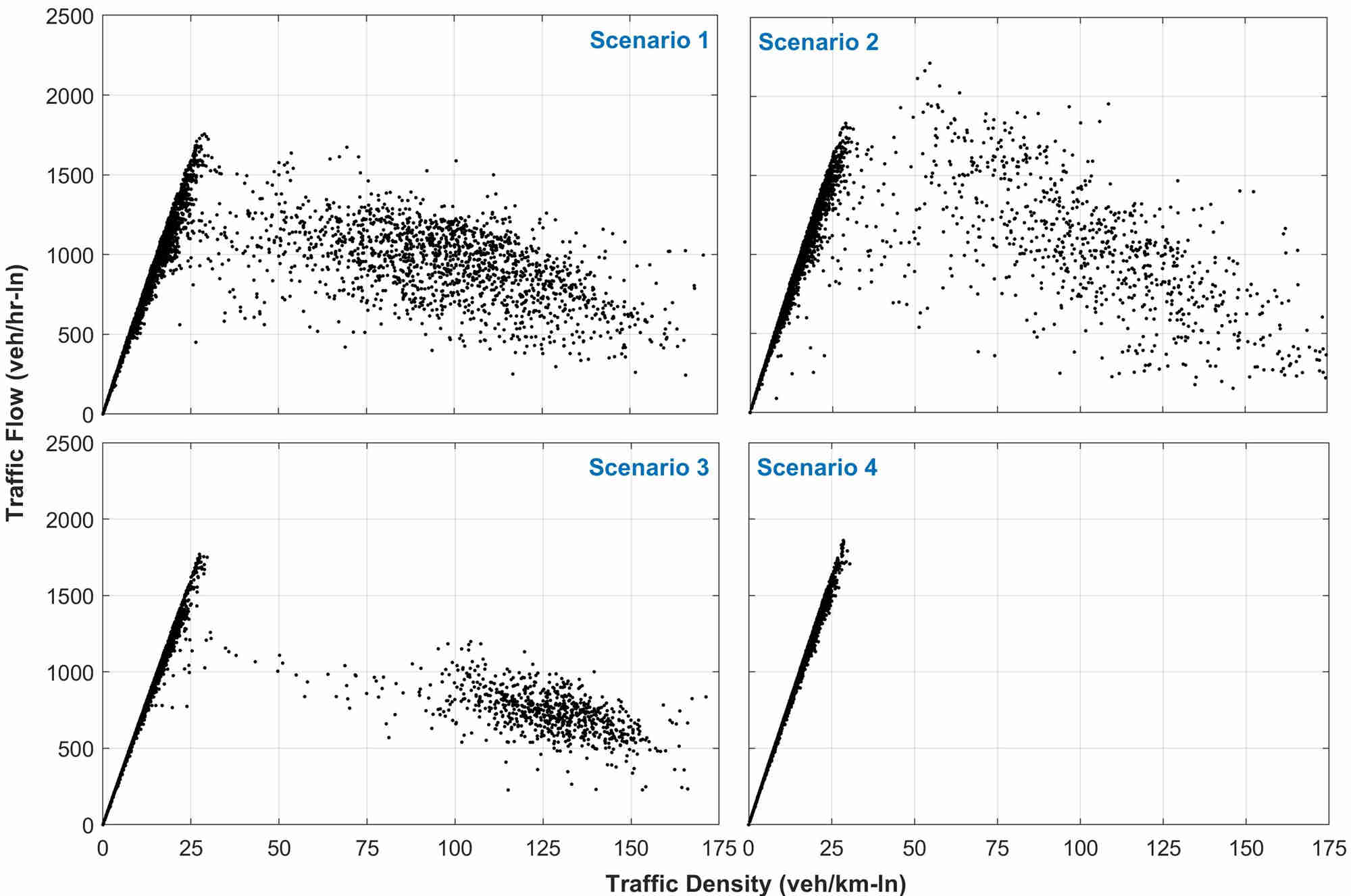}
\caption{Eastbound}
\label{fig:eastbound}
\end{subfigure}	
\begin{subfigure}[b]{0.45\textwidth}
\includegraphics[width=\textwidth]{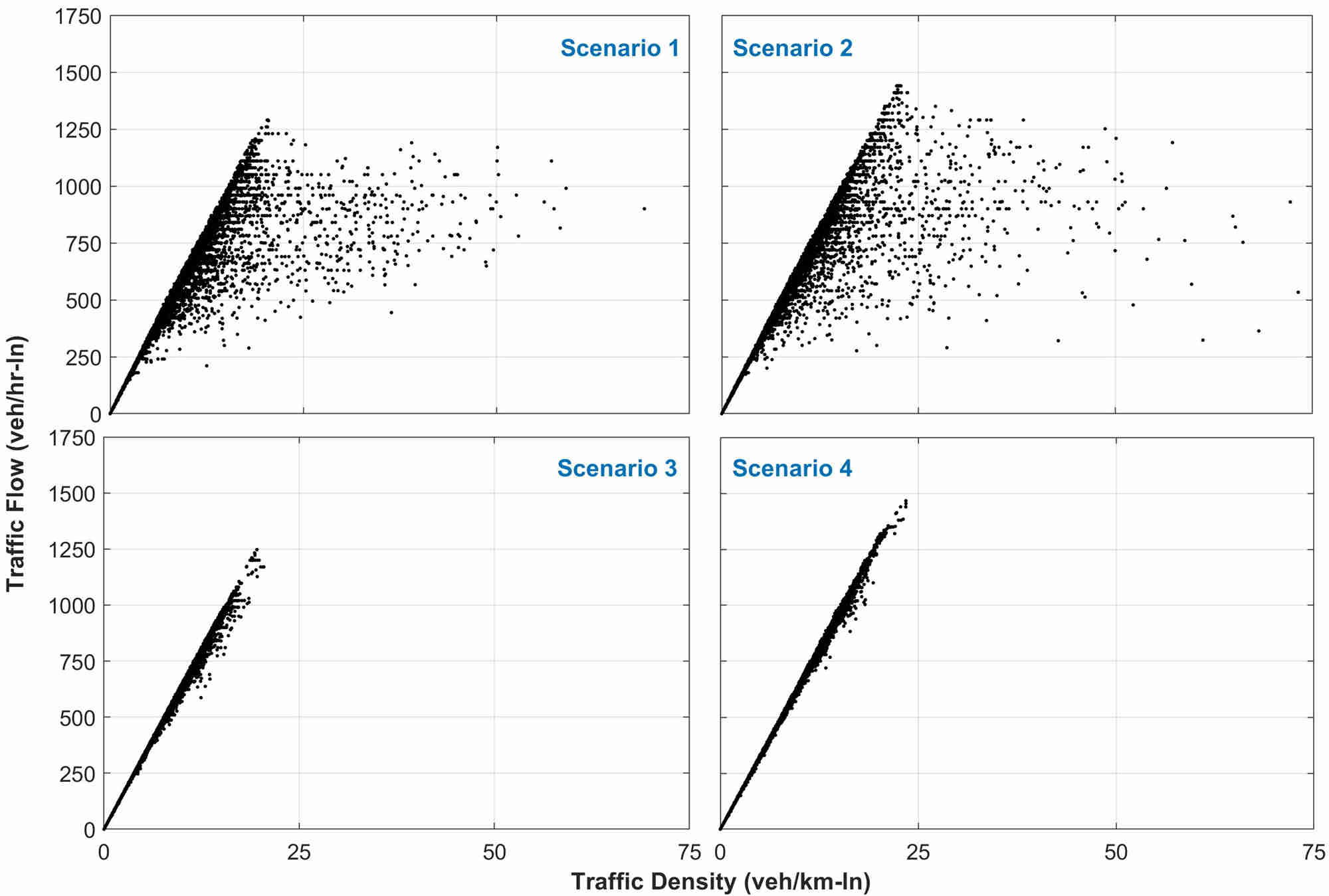}
\caption{Westbound}
\label{fig:westbound}
\end{subfigure}	

\begin{subfigure}[b]{0.45\textwidth}
\includegraphics[width=\textwidth]{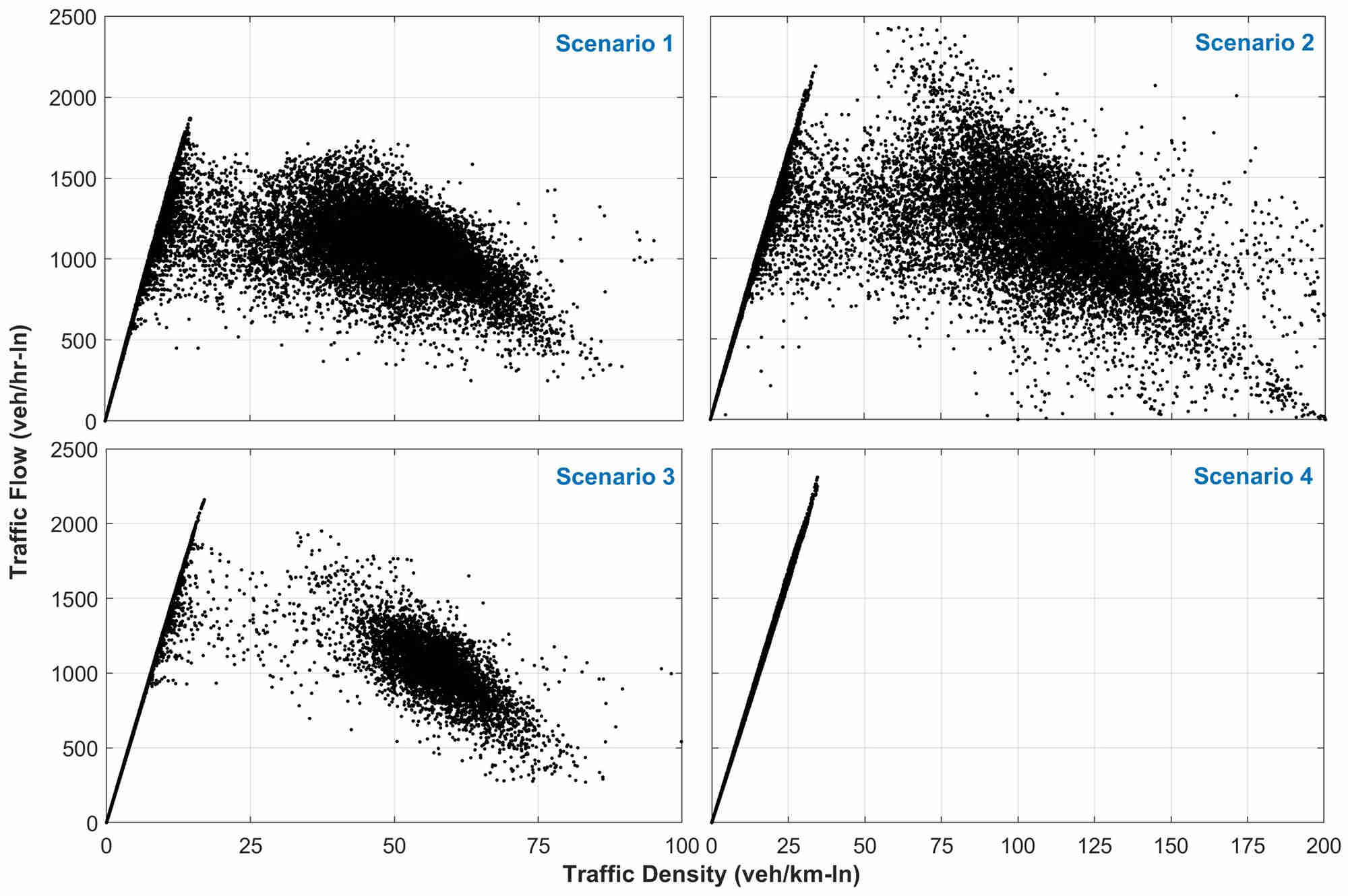}
\caption{Northbound}
\label{fig:northbound}
\end{subfigure}	
\begin{subfigure}[b]{0.45\textwidth}
\includegraphics[width=\textwidth]{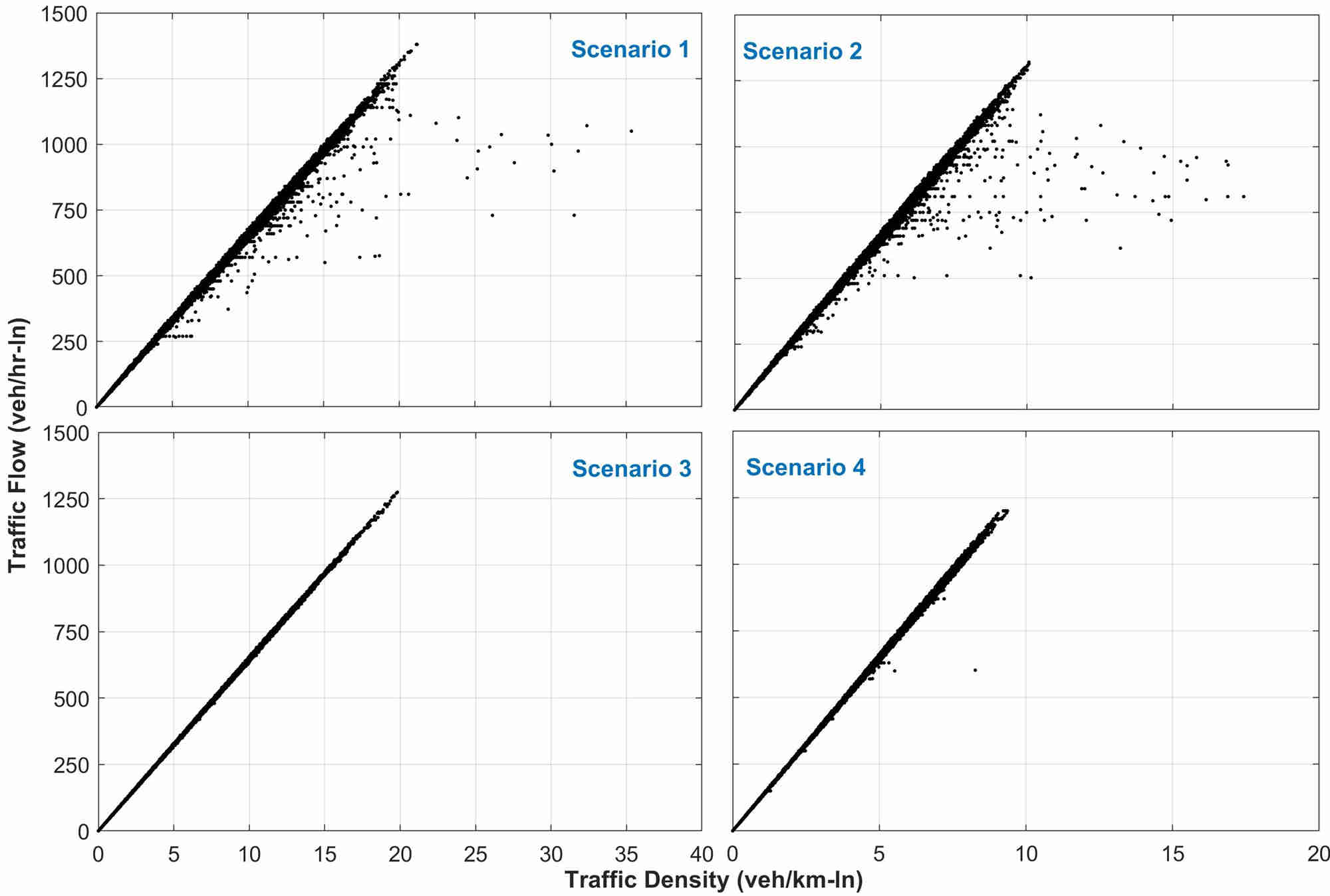}
\caption{Southbound}
\label{fig:southbound}
\end{subfigure}
\caption{Traffic flow vs. density.}
\label{fig:flow-density}
\end{figure*}

The relationship between traffic flow (vehicles per hour per lane, $veh/hr-ln$) and traffic density (vehicles per km per lane, $veh/km-ln$) under different scenarios is shown in Fig. \ref{fig:flow-density}. The eastbound and westbound (Fig. \ref{fig:eastbound} and \ref{fig:westbound}) represent local traffic between intersections 1 and 2, whereas the northbound and southbound (Fig. \ref{fig:northbound} and \ref{fig:southbound}) represent local to highway traffic between intersection 1 and the interchange. From Fig. \ref{fig:flow-density}, we can see that when the demand increases, the traffic gets congested in the network of uncontrolled human-driven vehicles and AVs (scenarios 1 and 2). Especially, when cycle lengths of two traffic signals are not sufficient enough to accommodate the increased traffic demand, the queues build up easily upstream the intersections and are difficult to resolve, e.g., traffic breakdown on northbound. 

Under scenario 3, traffic flow is generally improved with 1.2 $s$ headway setting, e.g., traffic congestion is resolved for westbound and southbound directions. However, when the traffic demand is extremely high, e.g., over 2000 $veh/hr-ln$ on the northbound, it is apparent that 1.2 $s$ headway and/or 150 $m$ control zone length limit the extent to which network capacity could be utilized such that all the traffic demand could be satisfied. Therefore, we observe that the optimal control fails in the extreme traffic demand cases, resulting in unresolved traffic congestion, i.e., eastbound and northbound traffic under scenario 3 in Fig. \ref{fig:flow-density}, and the increased travel time (Fig. \ref{fig:throughput}). If, however, we take into account the capability of shorter headways for the CAVs, the optimally controlled CAVs are able to form a smooth traffic flow and eliminate congestion in the network even with the highest traffic demand level (as shown by the linear flow-density relationship under scenario 4 in Fig. \ref{fig:flow-density}).

\section{Concluding Remarks}
In this paper, we compared the performance of a transportation network between optimally controlled CAVs and uncontrolled human-driven vehicles. The results highlight the impact of coordination of CAVs in terms of fuel economy and traffic safety. We have shown that the control parameters of the decentralized framework need to be handled carefully to fully utilize roadway capacity and satisfy different traffic demand levels in the network.
While the potential benefits of full penetration of CAVs to alleviate traffic congestion and reduce fuel consumption have become apparent, different penetrations of CAVs can alter significantly the efficiency of the entire system. Future work should investigate the implications of different penetration CAV penetration rates.


\bibliographystyle{IEEEtran}
\bibliography{ccta_traffic_ref}

\end{document}